# Experimental study on determinants of the evacuation performance in the super-high rise building

Fang Zhiming[1,*], Gao Huisheng[1], Huang Zhongyi[1], Ye Rui[1,**], Li Xiaolian[2], Xu Qingfeng[3], Xu Xingpeng[1], Huang Nan[1]

[1] Business School, University of Shanghai for Science and Technology, Shanghai 200093, China

[2] Colleage of Ocean Science and Engineering, Shanghai Maritime University, Shanghai 201306, China

[3] Shanghai Key Laboratory of Engineering Structure Safety, Shanghai Research Institute of Building Science, Shanghai 200032, China

* Corresponding author. E-mail: zhmfang2015@163.com

** Corresponding author. E-mail: yerui@mail.ustc.edu.cn

**Abstract:** The stairwell is the main path for emergency evacuation of people in super high-rise buildings, so revealing the movement characteristics in the stairwell based on experimental data is the basis for controlling the evacuation process of super high-rise buildings to ensure the safety of the crowd. Here, an evacuation experiment is carried out in Shanghai Tower with a vertical height of 583 m, which is the second tallest building in the world. The results show that pedestrians would set a "suitable velocity" for themselves according to the target height distance, and farther distance will result in lower "suitable velocity". The evidence is that the group of participants with a 9.63% higher traveling height spend a 16.39% longer evacuation time, yet within each group the velocity do not decrease with the increase of the moving distance. Furthermore, crowding in stairwell determines whether pedestrians can achieve the "suitable velocity", and the "suitable velocity" of women or older people is smaller than men or younger people in the same scenario. The local velocity and vertical velocity of different groups, genders and ages are classified and analyzed. A new measurement method for crowd density and then the fundamental diagram of the velocity-density relation in super high-rise building is presented. These results can provide basic data for the design of emergency evacuation facilities and formulation of emergency plan for super high-rise buildings.

**Introduction**

The emergency evacuation process of super high-rise buildings involves large-scale crowds, multiple floors, and multiple evacuation channels, which is a typical complex system. In recent years, researchers have employed various scientific methods to study the evacuation characteristics of super high-rise buildings to explore methods to improve the evacuation efficiency of super high-rise buildings. Studies have shown that longitudinal evacuation will occupy more than 70% of the total emergency evacuation time of super high-rise buildings. Therefore, stairwell evacuation is the most critical part during the process of crowd evacuation in super high-rise buildings[1].

In response to the evacuation process in the staircase of super high-rise and high-rise buildings, in recent years, researchers had organized various experiments to extract and analyze the crucial parameters and influencing factors of emergency evacuation, including basic parameters that can characterize the evacuation efficiency of super high-rise and high-rise buildings, i.e. speed of walking downstairs[2-6], the relation between evacuation speed and crowd density in stairwell[2, 4, 6, 7], the speed of walking upstairs in the scenario that move towards to the nearest refuge floor above[5, 8, 9], the difference of evacuation performance in stairwells of pedestrians with different gender, age and weight[9], the effect of the group phenomenon because of family and friends on crowd flow in stairwells [10, 11], and the influence of light condition, visibility, and other environmental factors on the evacuation process on stairs[12].

Regarding the characteristics of the movement on the stairs, the relevant experimental results showed that: 1) compared with horizontal movements, there would be obvious queuing when pedestrians gone downstairs[2, 7, 11]; 2) when the crowd on each floor met the crowd in the staircase (referred to as the confluence process), the crowd on the floor would dominate[13], and the confluence would reduce the evacuation efficiency of the crowd in the staircase[6, 11, 14, 15]; 3) it was easy to form small groups (a group with two to three people) in the stairwell. The speed of the group was often smaller than the speed of a single pedestrian, which would affect the movement speed of other people[10, 11]; 4) the overall speed in the stairwell was more likely to be affected by individual with lower speed than on the regular floor [2, 11]; 5) the width of stairs had a great influence on the pedestrian flow rate[11, 14]; 6) the decrease in visibility had a significant effect on the movement process on the stairs[12].

For the evacuation process in the stairwell with long-distance stairs of super high-rise buildings, the experimental results revealed that: 1) decreased physical strength might have an impact on the evacuation process. For instance, in the 50-story evacuation experiment[5], there was no obvious deceleration process when pedestrians moved downward, but there was an obvious deceleration behavior after pedestrians moved upward 20 floors. In the evacuation experiment of a 101-story building[3], the downward speed of personnel decreased significantly on the last 6 floors; 2) compared with women, men presented an obvious advantage in athletic ability during long-distance evacuation, which was manifested in that the average evacuation speed of men was significantly higher than that of women[5, 9]; 3) the evacuation speed had a certain correlation with age[9], but not with weight[5].

Table1 An overview of the evacuation experiments in high-rise buildings

| Authors and year | Participant number | Floor number | Focus |
| --- | --- | --- | --- |
| Fang et al.(2012) [2] | 163 | 8 | Downward velocity, velocity-density relation |
| Ma et al. (2012) [3] | 6 | 101 | Downward velocity |
| Peacock et al.(2012) [4] | 525/593/1148 226/127/793 | 31/24/18 13/11/10 | Downward velocity, velocity-density relation |
| Choi et al.(2014) [5] | 60 | 50 | Downward velocity, upward velocity |
| Lam et al.(2014) [9] | 120 | 40 | Downward velocity, kinds of influence factors |
| Huo et al.(2016) [6] | 73 | 9 | Downward velocity, velocity-density relation |
| Zeng et al.(2017) [12] | 39 | 9 | Effect of visibility on velocity |
| Chen et al.(2017) [8] | 165 | 20 | Upward velocity |
| Ma et al. (2017) [10] | 56 | 11 | Effect of small group on velocity |

With the development of computer simulation technology, modeling and simulating the evacuation process of people in buildings had become an auxiliary method for analyzing evacuation problems. Considering the particularity of the space structure and evacuation path in the staircase, the researchers constructed a variety of staircase evacuation models, including using a finer grid to describe the structure of the staircase, which could accurately describe the evacuation environment of the staircase and the personnel motion restricted by the steps [16]; introducing the collection rate parameter into the model to study the impact of the horizontal and vertical crowd confluence on the evacuation process of the staircase[17]; establishing corresponding rules in the model to consider the impact of personnel's physical and psychological conditions[18] and crowd stress[19] on the staircase evacuation process; setting up a simulated person with limited mobility in the model and considering its impact on the entire staircase evacuation process[20]. These models could simulate the process of pedestrians' rotation and downward movement in the stairwell[16, 18, 20, 21] and the process of confluence[16, 17], and reproduce the special phenomena of pedestrians waiting in line[19, 20], preferring the inner path[18] and layering within people flow[21, 22].

The verified evacuation model was used as a tool to analyze the influencing factors on the evacuation efficiency of super high-rise buildings. The related simulation results showed that the crowd confluence process would form a crowded state at the floor platform near the entrance and exit of the staircase, thereby reducing the crowd evacuation speed[16, 21]. In the process of converge, if the entrance of the staircase was on the side of the lower staircase, the crowd movement in the staircase would dominate, and the entrance of the staircase on the side of the upper staircase would facilitate the entrance of the crowd on the staircase[23]. As the number of people on the floor platform increased, the evacuation time showed an approximately linear increase trend[21]. For the overall evacuation scene of super high-rise buildings, increasing the speed of personnel could not reduce the evacuation time without limit[24]. People with limited mobility would seriously reduce the overall evacuation efficiency, so it was necessary to focus on the emergency evacuation process of these people[20].

Therefore, experiment and modeling methods are currently effective research methods for the evacuation law of super high-rise buildings. Experiments can reveal the basic movement parameters of personnel in the evacuation process of super high-rise buildings, which is the basis to build and validate the evacuation model. The verified evacuation model could be used to analyze the evacuation problem of super high-rise buildings under various working conditions. However, there are only a little study on the evacuation process and results of super-high rise buildings. A vertical evacuation experiment in a 126-story (580m) super high-rise building is carried out in this study. The effects of gender, age, congestion and evacuation distance of individuals on the evacuation process are focused on. The remainder is organized as follows: Section 1 presents the experiment conducted in a super high-rise building. Section 2 gives analysis results about some key crowd movement characteristics in stairwells. The results are compared with previous studies and discussed in Section 3, and the summary of our work is in Section 4.

## 1 Methods and data collection

### 1.1 Experiment

The experiment is carried out in Shanghai Tower (Shanghai, China), a 632 m high building with 126 stories. It is the second tallest building in the world, and the tallest building in China. There are four stairwells in this building, one of which is chosen to conduct the experiment. This selected stairwell connects the 126$^{th}$ floor to the ground. The width, riser and tread of most stairs are about 1.2 m, 0.155 m and 0.25 m respectively. Totally 90 participants are recruited, whose age ranged from 20 to 51 years old. The experiment is conducted in the morning (no rain, bright), and the movement of each participant is obtained from the surveillance cameras on certain floors distributed on the top of stair landing.

Participants are divided into three groups according to the floor from which they start to evacuate. More specifically, Group 1 with 44 participants evacuate from the 126$^{th}$ floor (the relative height is 583.43m), Group 2 with 30 participants evacuate from the 117$^{th}$ floor (the relative height is 542.10 m) and Group 3 with last 16 participants evacuate from the 100$^{th}$ floor (the relative height is 465.50 m), respectively. The participants in Group 1 and Group 2 are recruited from the community, and the participants in Group 3 are all security guards of the building. The instructions given to the participants are as follows:

- All participants should walk downstairs to the ground (the relative height is 0) with their preferred speeds.
- The participants in Group 1 from the 126$^{th}$ story start to move firstly.
- When the foremost of Group 1 arrived at the 117$^{th}$ story, participants in Group 2 start to move.
- When the foremost of Group 1 or Group 2 arrived at the 100$^{th}$ story, participants in Group 3 start to move.

### 1.2 Performance

Most of the participants in Group 3 evacuate by elevator in some floors unexpectedly for that they have the access control card to the elevator. 3 participants in Group 1 and 2 in Group 2 quit midway for the reason of fatigue or health problem. The position of the stairwell changes on the 7$^{th}$ floor but without clear instruction of the new position, which causes that most of participants are lost in the corridor of 7$^{th}$ floor for a moment. As a result, only the process of 69 participants (41 in Group 1 and

28 in Group 2, as shown in Table 2 and Fig. 1.) moving from their initial floor to the 7th floor (the relative height is 33.45 m) are considered in this study.

Table 2 Age and gender distribution of 69 participants considered.

|  |  | Group 1 | Group 2 |
|---|---|---|---|
| **Gender** | Male | 26 | 23 |
|  | Female | 15 | 5 |
| **Age** | 20-29 | 20 | 21 |
|  | 30-39 | 16 | 7 |
|  | 40-50 | 3 | 0 |
|  | 50-60 | 2 | 0 |
| **Total** |  | 41 | 28 |

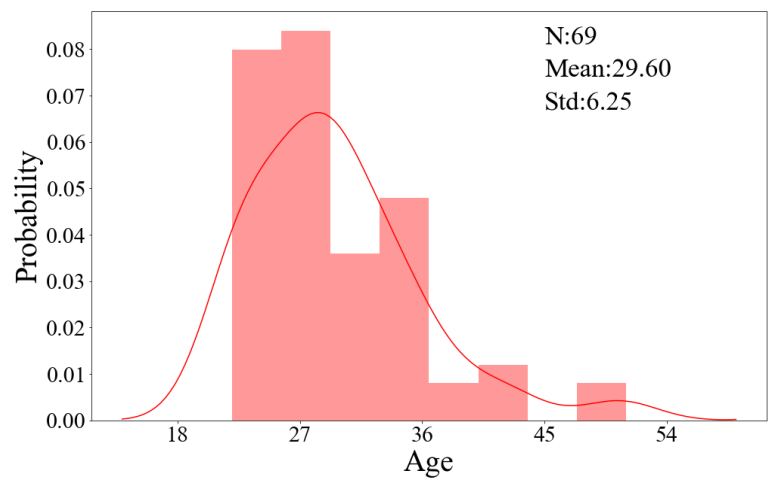

Fig. 1 Age distribution of 69 participants considered.

**1.2 Stair sections**

It is difficult to consider the movement characteristics in one stair section after another, for the reason that not all floors are equipped with surveillance camera. For example, there is only one camera installed on the 99th floor between the 84th and 105th floor. Furthermore, each stair section between two adjacent floors may include 2 or 4 flights of stairs. For consistency, this study only focus on total 58 monitored stair sections, including 56 stair sections between two adjacent floors with 2 flights of stairs and 2 long stair sections (S106-99 and S99-93), as shown in Table 3. The 2 long stair sections are used to ensure that there are data in each 10 floors. Furthermore, S125-7 and S115-7 are used to represent the overall stair sections of the two groups respectively.

Table 3 Stair sections considered in this study

| ID | label | $n_s$ | $n_f$ | $l_t$ | ID | label | $n_s$ | $n_f$ | $l_t$ |
|---|---|---|---|---|---|---|---|---|---|
| 1 | S125-124 | 28 | 2 | 2.6 | 31 | S48-47 | 28 | 2 | 0 |
| 2 | S124-123 | 25 | 2 | 0.25 | 32 | S47-46 | 28 | 2 | 0 |
| 3 | S123-122 | 25 | 2 | 0.25 | 33 | S46-45 | 28 | 2 | 0 |
| 4 | S122-121 | 25 | 2 | 1 | 34 | S45-44 | 28 | 2 | 0 |
| 5 | S115-114 | 26 | 2 | 0 | 35 | S44-43 | 28 | 2 | 0 |
| 6 | S114-113 | 26 | 2 | 0 | 36 | S43-42 | 28 | 2 | 0 |
| 7 | S113-112 | 26 | 2 | 0 | 37 | S42-41 | 28 | 2 | 0 |
| 8 | S112-111 | 26 | 2 | 0 | 38 | S41-40 | 28 | 2 | 0 |
| 9 | S111-110 | 26 | 2 | 0 | 39 | S40-39 | 28 | 2 | 0 |
| 10 | S110-109 | 26 | 2 | 0 | 40 | S39-38 | 28 | 2 | 1.25 |
| 11 | S109-108 | 26 | 2 | 0 | 41 | S33-32 | 28 | 2 | 0 |
| 12 | S108-107 | 26 | 2 | 0 | 42 | S32-31 | 28 | 2 | 0 |
| 13 | S106-99 | 226 | 20 | 3.75 | 43 | S31-30 | 28 | 2 | 0 |
| 14 | S99-83 | 432 | 36 | 4.75 | 44 | S30-29 | 28 | 2 | 0 |
| 15 | S78-77 | 28 | 2 | 0 | 45 | S29-28 | 28 | 2 | 0 |
| 16 | S77-76 | 28 | 2 | 0 | 46 | S28-27 | 28 | 2 | 0 |
| 17 | S76-75 | 28 | 2 | 0 | 47 | S27-26 | 28 | 2 | 0 |
| 18 | S75-74 | 28 | 2 | 0 | 48 | S26-25 | 28 | 2 | 0 |
| 19 | S74-73 | 28 | 2 | 0 | 49 | S25-24 | 28 | 2 | 0 |
| 20 | S71-70 | 28 | 2 | 0 | 50 | S24-23 | 28 | 2 | 0 |
| 21 | S70-69 | 28 | 2 | 0 | 51 | S18-17 | 28 | 2 | 0 |
| 22 | S65-64 | 28 | 2 | 0 | 52 | S17-16 | 28 | 2 | 0 |
| 23 | S64-63 | 28 | 2 | 0 | 53 | S16-15 | 28 | 2 | 0 |
| 24 | S63-62 | 28 | 2 | 0 | 54 | S15-14 | 28 | 2 | 0 |
| 25 | S62-61 | 28 | 2 | 0 | 55 | S14-13 | 28 | 2 | 0 |
| 26 | S58-57 | 28 | 2 | 0 | 56 | S13-12 | 28 | 2 | 0 |
| 27 | S57-56 | 28 | 2 | 0 | 57 | S10-9 | 28 | 2 | 0 |
| 28 | S56-55 | 28 | 2 | 0 | 58 | S8-7 | 28 | 2 | 1.25 |
| 29 | S55-54 | 28 | 2 | 0 |  | S125-7 | Overall section of Group1 | | |
| 30 | S54-53 | 28 | 2 | 1.25 |  | S115-7 | Overall section of Group2 | | |

## 1.3 Data collection

### *a. Movement time*

Detailed movement process of each participant in the stairwell have been recorded by cameras. As shown in Fig. 2, the process of a typical male participant reaching and then leaving the stair section S63-62 is recorded by the camera on the 63[th] and 62[th] floor. The reaching time, which means the moment when each participant reaches the landing above of each stair section, can be extracted manually, and then the movement time that each participant spent ($\Delta t$) in each stair section can be obtained by calculating the difference of reaching times between two adjacent stair sections. Furthermore, the time spent in the two overall stair sections are labeled using $\Delta T_{125-7}$ and $\Delta T_{115-7}$ respectively.

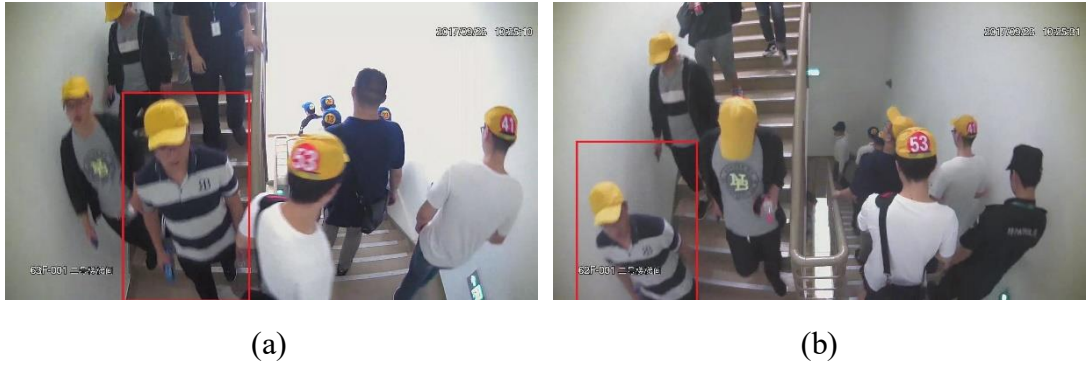

(a)          (b)

Fig. 2 The reached moment of a participant on the landing of (a) the 63$^{th}$ floor and (b) the 62$^{th}$ floor.

*b. Movement distance*

The movement distance in each stair section ($\Delta d$) is calculated as follows:

$$\Delta d = n_s\sqrt{s_t^2 + s_r^2} + n_f l_f + l_t \qquad (1)$$

where $s_t$ is the value of the stair tread (0.25 m), $s_r$ is the value of the stair riser (0.155 m), $n_s$ is the number of steps in each stair section. $n_f$ is the number of turn-back that pedestrians need in each stair section. Depending on the structure of stairwell, pedestrians may turn 180° or 90° most of the time on the landing, and then $n_f$ adds 1 for 180° or 0.5 for 90°. $l_f$ is the distance of a turn-back. For simplicity, we assume that participants would walk along a semicircle or a quarter-circle with a radius of 0.3m on the landings, so $l_f = 0.3\pi n_f = 0.942$ m. Besides, $l_t$ is the distance of the non-normal plane that connecting two stair sections in some floors. The detailed value of $n_s$, $n_f$, and $l_t$ are shown in Table 3. Furthermore, the movement distance in the two overall stair sections are labeled using $\Delta D_{125-7}$ (1356.76 m) and $\Delta D_{115-7}$ (1251.94 m) respectively.

*c. velocity*

The average velocity of each participant walking through each stair section is defined as local velocity (labeled using $v_{Si-j}$), which can be calculated as follows:

$$v_{Si-j} = \Delta d/\Delta t \qquad (2)$$

The average velocity of each participant walking through the two overall stair sections are defined as overall velocity (labeled using $V_{125-7}$ and $V_{125-7}$), which are calculated as follows:

$$V_{125-7} = \Delta D_{125-7}/\Delta T_{125-7}; \ V_{115-7} = \Delta D_{115-7}/\Delta T_{115-7} \quad (3)$$

The height of each participant walking down per unit time in the two overall stair sections are defined as overall vertical velocity (labeled using $V^h_{125-7}$ and $V^h_{115-7}$), which are calculated as follows:

$$V^h_{125-7} = \Delta H_{125-7}/\Delta T_{125-7}; \ V^h_{115-7} = \Delta H_{115-7}/\Delta T_{115-7} \quad (4)$$

where $\Delta H_{125-7}$ is the height difference between the 125$^{th}$ and 7$^{th}$ floor (545.95 m), and $\Delta H_{115-7}$ is the height difference between the 115$^{th}$ and 7$^{th}$ floor (497.95 m)

*d. density*

The traditional method for the calculation of crowd density is difficult to utilize in the stairwell of super high-rise buildings. The method from the single file movement to calculate density is adopted in this study. According to the Ref. [25, 26], pedestrians start to walk with their preferred/desired speed (1.29 m/s) when the distance headway reaches 3 m, which corresponds to a time interval of 2.326 s (3/1.29). We then measure the density ($\rho$) in the stair section with the help of this time interval. More specifically, let the time when one pedestrian reaches one stair section denote as $t_0$, then the time range 2.326 s before ($t_{-1}$) and after ($t_{+1}$) it is defined in Eq. (5).

$$t_{-1} = t_0 - 2.326; \ t_{+1} = t_0 + 2.326 \quad (5)$$

The number of pedestrians ($N_p$) who reach this stair section during the above-mentioned time range ($t_{-1} \sim t_{+1}$) can be calculated, which can be viewed as a flow number in a certain time interval as shown in Fig. 3. Finally, according to the basic traffic flow relation (Flow = Density·Velocity), the density ($\rho$) in selected stair sections can be calculated as follows:

$$\rho = {N_p} \Big/ {\overline{v_{Si-j}} \cdot \sigma_t \cdot W_S} \quad (6)$$

Where $W_S$ is the stair width (1.20 m), $\sigma_t = t_{+1} - t_{-1}$ is the time interval (4.652 s), and $\overline{v_{Si-j}}$ is the average value of the local velocity in the corresponding stair section of the $N_p$ pedestrians.

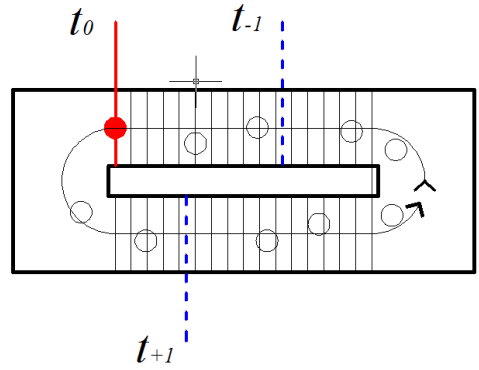

Fig. 3 Flow number in a certain time interval around the reaching time of the pedestrian represented by a filled circle (the red solid line represents the moment when this pedestrian reaches the stair section ($t_0$), and the time 2.326s away from this moment ($t_{-1}$, $t_{+1}$) are represented using two blue dot lines. The 5 pedestrians within the two dot line reach the stair section during the time range ($t_{-1} \sim t_{+1}$), so $N_p$ is 5).

## 2 Results

### 2.1 Comparison of the two groups

Fig.5 shows the spatial-time diagram of the evacuation process of the two groups. The blank areas mean that there is no camera on the corresponding floor. As pedestrians evacuate downstairs, the difference in evacuation time for individuals on the same floor increase gradually, which results from the heterogeneity of walking speeds among participants. Finally, the overall evacuation time $\Delta T_{125-7}$ of Group 1 is 33.80±6.02min, and $\Delta T_{115-7}$ of Group 2 is 29.04±5.62min. An increase by 9.63% of the height causes the evacuation time increases by 16.39%. Moreover, $\Delta T_{115-7}$ of Group 1 is 31.16±4.78 min and obviously more than $\Delta T_{115-7}$ of Group 2.

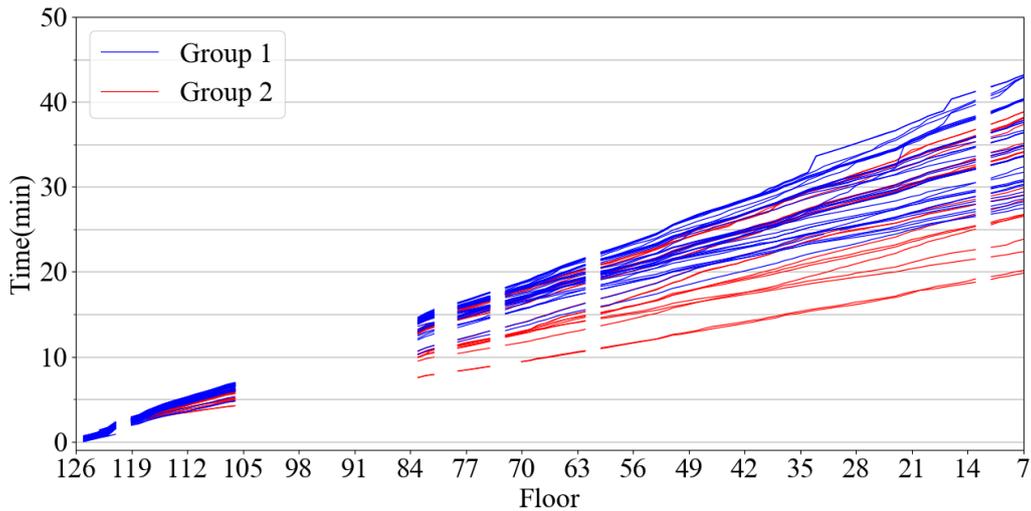

Fig.4 Spatial-time diagram of the evacuation process for the two groups.

The local velocity $v_{Si-j}$ of all the 69 participants in the 58 stair sections are calculated according to Eq. (2), and then classified by groups. Fig. 5 shows that the local velocity of the two groups basically conform to the normal distribution. The mean value in Group 2 (0.767 ± 0.217 m/s) is larger than that of Group 1 (0.730 ± 0.213 m/s), which indicates that pedestrians in the super high-rise building may be more willing to evacuate with a lower speed as the height of their initial position increases. At the same time, the standard deviation in both groups is almost equal. The increase in evacuation distance doesn't make local velocity more scattered.

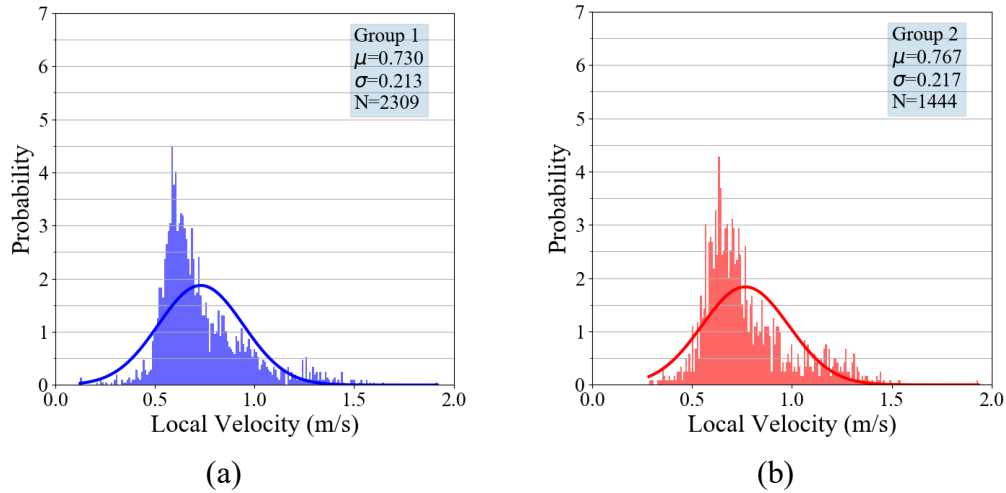

Fig. 5 Distribution of local velocity of two groups. (a): Group 1; (b): Group 2

To detect the effect of travelled height on the velocity, all the data points of local velocity in every 10-stories are extracted, of which the average value and distribution are shown in Fig. 6. For example, there are two stair sections (S10-9, S8-7, as shown in Table 3) below the 10$^{th}$ floor, and then $v_{S10-9}$ and $v_{S8-7}$ of Group 1 are regarded as one data set. The rightmost blue circle point in Fig. 6a is the average value, and the blue histogram at the bottom right corner of Fig. 6b (which is also right under the circle point in Fig. 6a) is the distribution of this data set of Group1. The curves in Fig. 6a show the variation tendency of local velocity as participants walk downstairs, and the following observations are made:

1) The local velocity shows a downward trend before the 100$^{th}$ floor especially in Group 1. This phenomenon may be caused mainly by two factors. On one hand, two times of confluence are set on 117$^{th}$ floor and 100$^{th}$ floor in this experiment, thus most of participants in Group 1 are blocked twice (by Group 2 and Group 3) and Group 2 are blocked once (by Group 3). On the other hand, participants have not dispersed sufficiently at the beginning.

2) The local velocity of Group 1 keeps an obvious increasing trend after the 100$^{th}$

floor especially when walking from 100$^{th}$ to 50$^{th}$ floor. As mentioned above, confluence and crowding result in lower velocity, which is not the desired velocity of participants, especially in Group 1. Therefore, once pedestrians in the stairwell present the trend of more and more decentralization as the evacuation experiment goes on, they will speed up to reach their desired velocity.

3) The average local velocity of Group1 in each 10-stories is always lower than Group 2. This may be caused by the reasons that participants in Group 1 need to evacuate from higher position, and there are more women and older people in Group 1 (see Table 2).

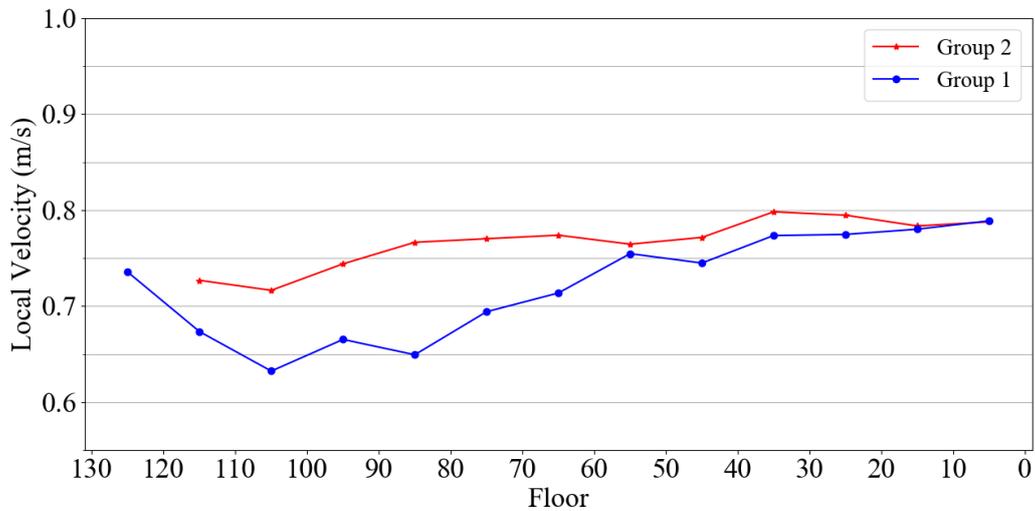

(a)

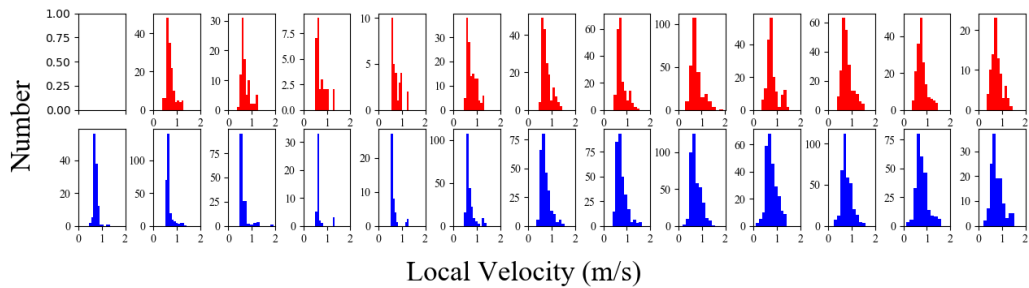

(b)

Fig. 6. The change of local velocity as the increase of evacuation distance for the two groups. (a): The average value of local velocity in each 10-stories; (b) The distribution of local velocity in each 10-stories

## 2.2 Influence of gender

To analyze the difference of escape ability between different genders in super high-rise building, all data of local velocity are classified by genders and groups (Fig. 7), and the change of local velocity of different genders as the increase of evacuation

distance is presented (Fig. 8). The results are as follows:

1) The average velocity of female (0.656 ± 0.141 m/s) is obviously lower than male (0.780 ± 0.229 m/s). Moreover, the velocity in each 10-stories of female is always lower than male, as shown in Fig. 8a.

2) Male velocity of Group 1 is a little lower than Group 2, but female velocity of Group 1 is much lower than Group 2, which indicates that women are more sensitive to the increase of the vertical evacuation distance.

3) As participants walk downstairs after the confluence, male velocity accelerates significantly and yet there is no obvious increase in female velocity.

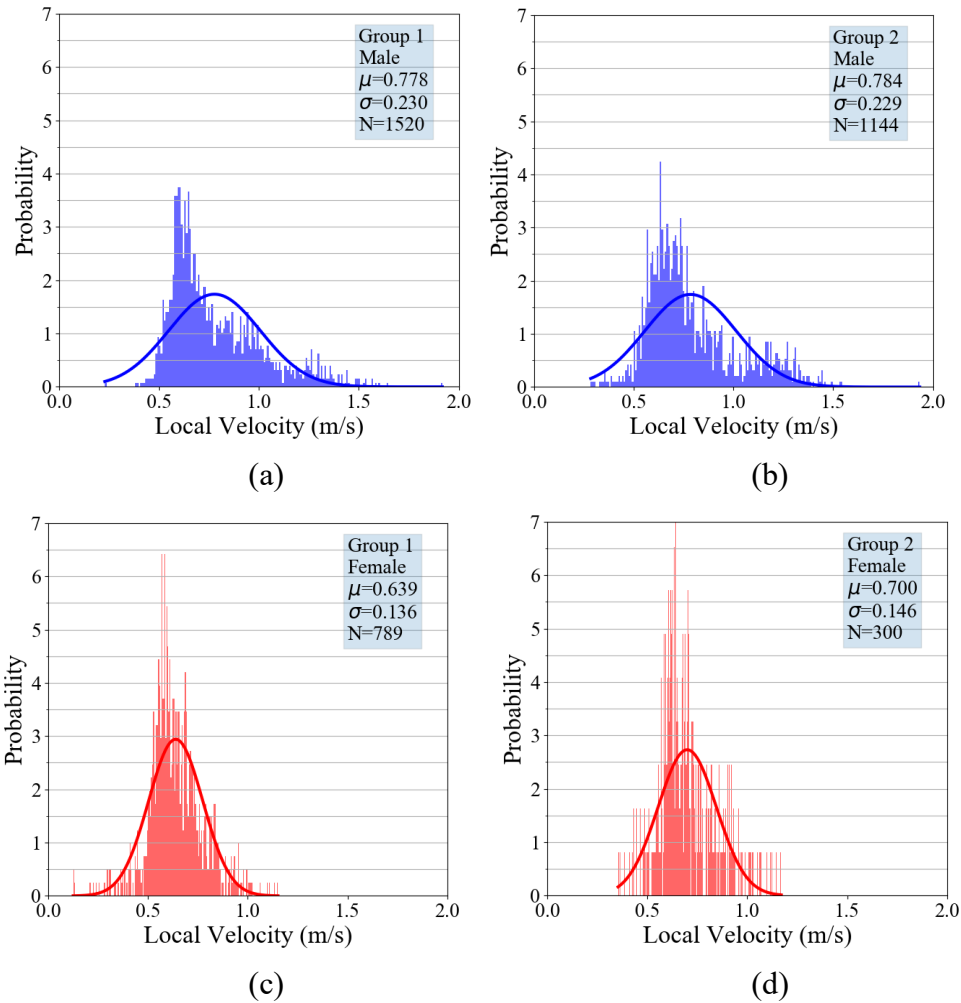

Fig. 7 Distribution of local velocity of different genders. (a): male of Group 1; (b): male of Group 2; (c): female of Group 1; (d): female of Group 2

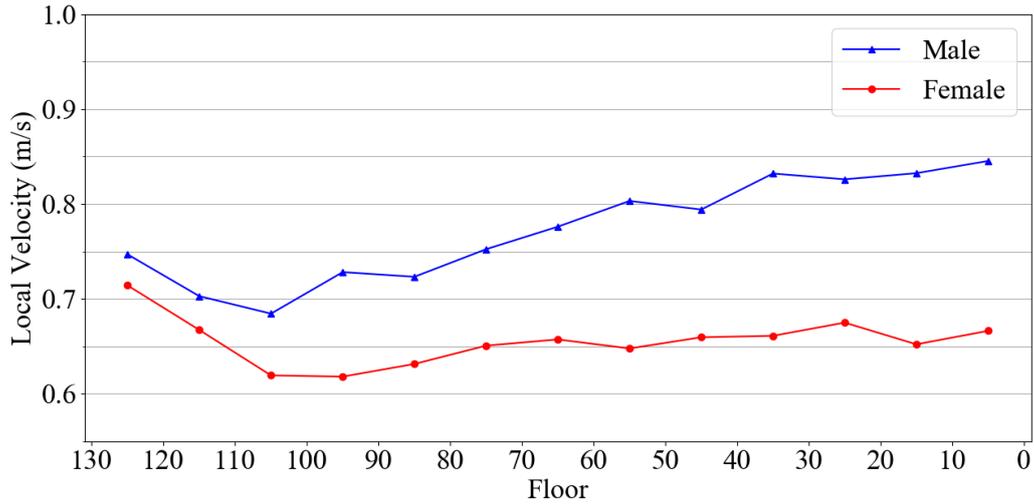

(a)

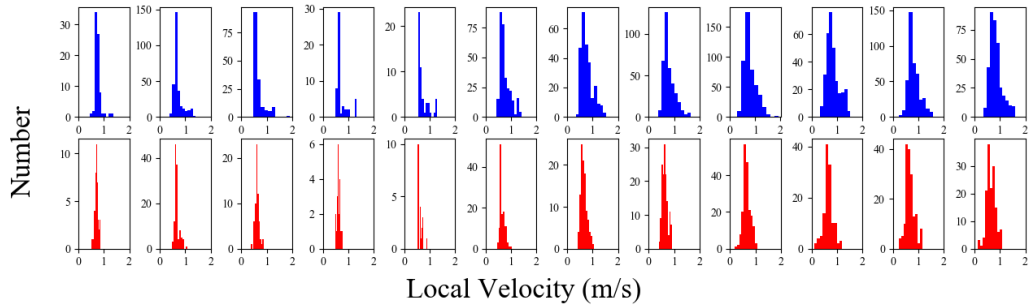

(b)

Fig. 8. The change of local velocity of different genders as the increase of evacuation distance. (a): The average value of local velocity in each 10-stories; (b) The distribution of local velocity in each 10-stories

**2.3 Influence of age**

To detect the effect of age on the escape ability in super high-rise building, all data of local velocity are classified by 2 age groups (<30 and ⩾30, Fig. 9), and the change of local velocity of different age groups as the increase of evacuation distance is presented (Fig. 10). The results are as follows:

1) The average velocity of younger people (age < 30) is 0.777 ± 0.229 m/s, and the average velocity of the older people (age ⩾ 30) is 0.699 ± 0.186 m/s. The younger people evacuate faster than the older people in super high-rise building, which is in line with the general perception.

2) Under the influence of the confluence, the velocity of both age groups shows a downward trend and then an upward trend as the increase of evacuation distance. However, the amplitude of decline and rise of younger people is much smaller than older people, which indicates that younger people are less affected by the confluence.

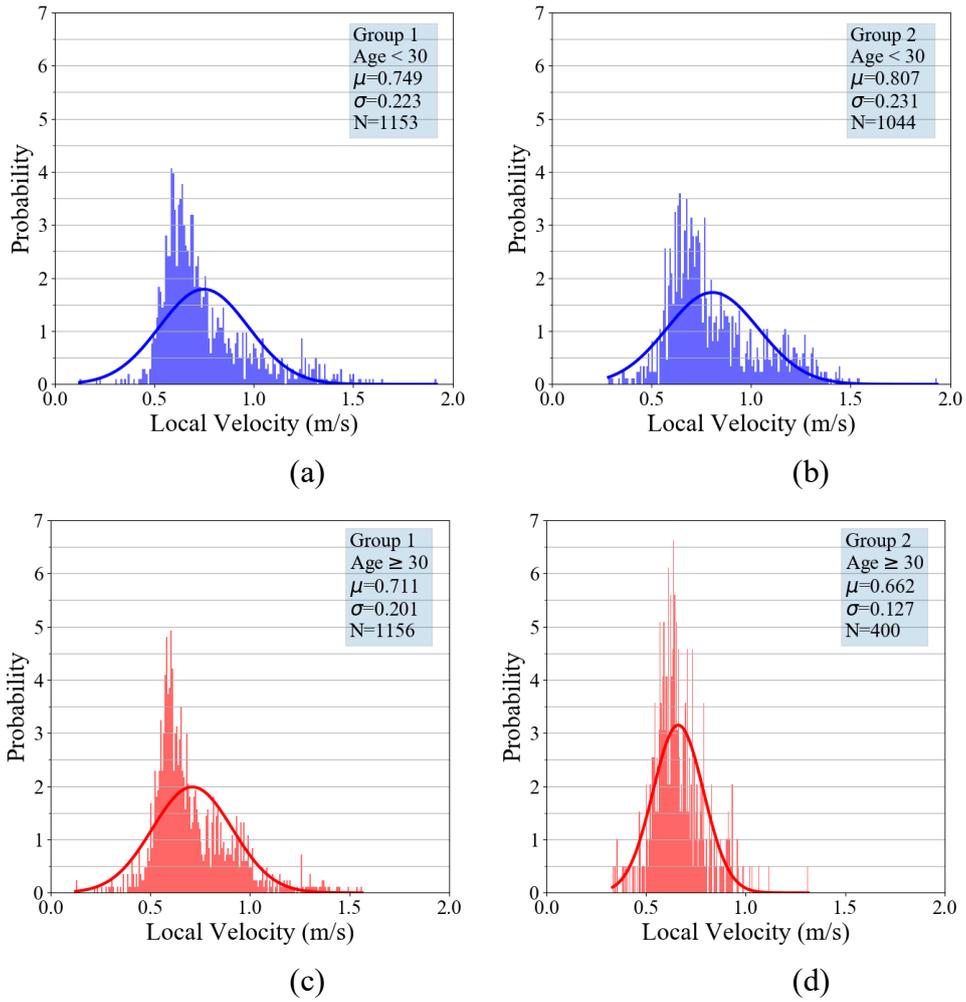

Fig. 9 Distribution of local velocity of different ages. (a): <30 of Group 1; (b): <30 of Group2; (c): ≥30 of Group 1; (d): ≥30 of Group2

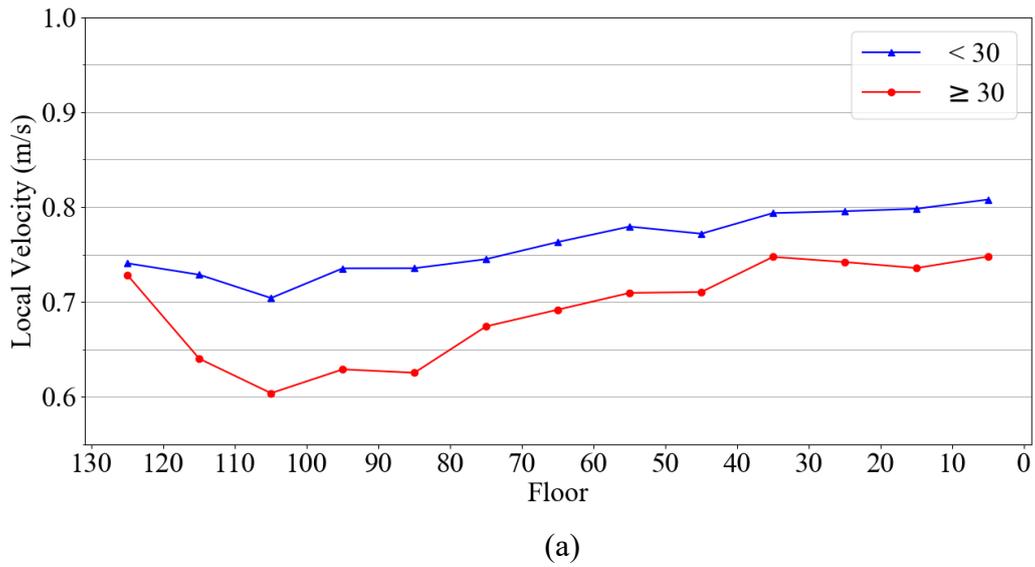

(a)

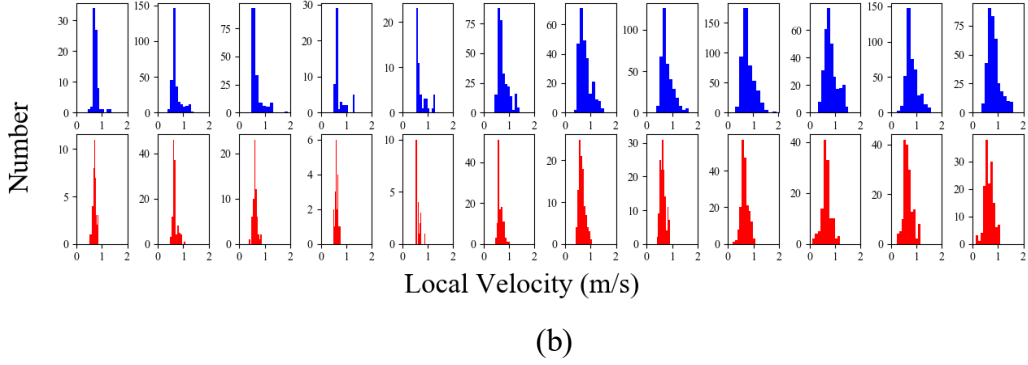

(b)

Fig. 10. The change of local velocity of different ages as the increase of evacuation distance. (a): The average value of local velocity in each 10-stories; (b) The distribution of local velocity in each 10-stories

**2.4 Relationship between density and velocity**

When evacuating through the stairwell, pedestrians from different regular floors will enter the stairwell at the same time, which will result in confluence and crowding. Although the number of participants is not large, two times of confluence are set on $117^{th}$ floor and $100^{th}$ floor in this experiment. Previous studies have presented pedestrians' velocity-density relationship in stairwell and indicated that the degree of congestion has a significant effect on the evacuation process[3, 27]. To detect the relationship between density and velocity in the evacuation process of super high-rise buildings, for each participant in each stair section, the data pair of density $\rho$ and local velocity ($v_{\text{Si}-j}$) are calculated by Eq. (6) and Eq. (2) respectively. As shown in Fig. 11, local velocity ($v_{\text{Si}-j}$) and density ($\rho$) basically show a negative correlation, and they satisfy as Eq. (7). As the density increases, the attenuation rate of local velocity gradually decreases.

$$v_{\text{Si}-j} = -0.189\rho + 0.863 \tag{7}$$

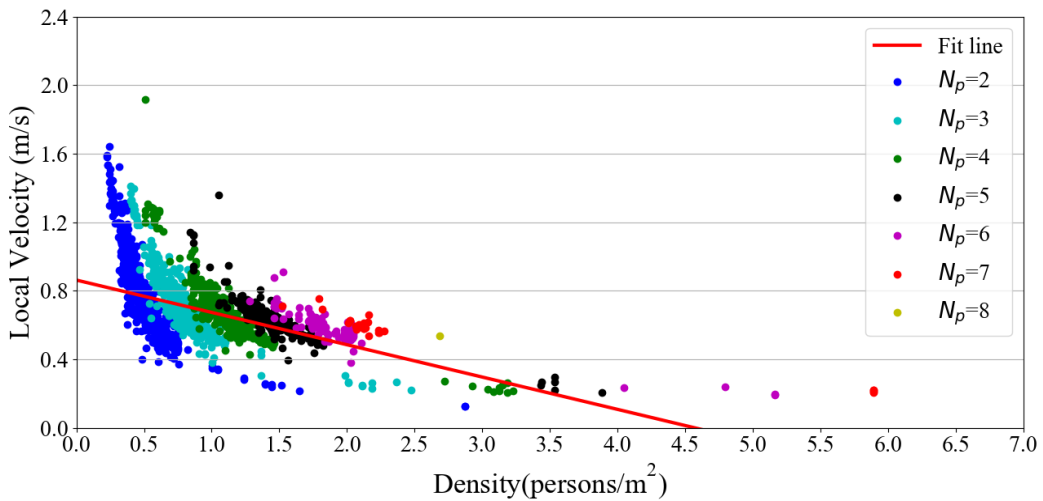

Fig. 11 The relationship between density and local velocity.

## 3 Discussion

### 3.1 Determinants of the evacuation velocity in super-high rise buildings

The local velocity (Eq. (2)), overall velocity (Eq. (3)), and overall vertical velocity (Eq. (4)) of different groups, genders and ages are calculated and listed in Table 4. Furthermore, the correlation analysis between the age, gender, group and the overall velocity $V_{115-7}$ are conducted, as shown in Table 4. The results show that the evacuation velocity in super high-rise building is negatively correlated with age and vertical distance, and related with gender. Summarizing the results of this experiment, the relationship between the evacuation velocity and distance, gender, age, and density in super high-rise building is as follows:

1) Distance: On the one hand, pedestrians' local velocity do not decrease with the increase of the moving distance (Fig. 6a). On the other hand, when walking through the same distance (for example, the stair section S115-7), the velocity ($V_{115-7}$, see Table 4) of Group 1 (of which the total vertical evacuation distance is larger) is significantly lower than that of Group 2. It seems that participants will set a "suitable velocity" for themselves according to the target distance (number of floors) before the evacuation, to ensure that they can evacuation with a relatively stable "suitable velocity" although their physical energy decline during the evacuation process. Moreover, the farther the target distance is, the lower the value of "suitable velocity" may be.

2) Density: A new measurement method for crowd density in super high-rise building is presented (Eq. (6)), and Fig. 11 shows that the local velocity tends to decrease as the density increases. Therefore, the density is a key factor in determining whether pedestrians can achieve the "suitable velocity". Specifically, pedestrians couldn't achieve their "suitable velocity" in high density condition, which manifests that the local velocity decreases when meeting confluence or crowding (Fig. 6a). Pedestrians will speed up to their "suitable velocity" in low density condition, which manifests that the local velocity increases after the confluence (Fig. 6a).

3) Gender: Females are significantly slower than males in the evacuation process of super high-rise building, thus female "suitable velocity" is smaller than male in the same scenario. Furthermore, the average velocity of women in Group 1 is 8.71% less than Group 2, but the average velocity of men in Group 2 is only 0.77% less than Group 2, which means that female "suitable velocity" is more sensitive to the change of the target distance. Moreover, as shown in Fig. 8a, the local velocity falls a lot under the influence of the confluence, but only increases a little after the confluence, which seems that female may reduce their "suitable velocity" after a period of lower velocity.

4) Age: Younger participants show better evacuation ability in super high-rise building, including that younger people evacuate faster than the older people, and younger people are less affected by the confluence.

Table 4 Velocity in the experiment (m/s)

|  |  | $v_{\text{Si-j}}$ | $V_{125-7}$ | $V_{115-7}$ | $V_{125-7}^h$ | $V_{115-7}^h$ |
|---|---|---|---|---|---|---|
| Group 1 | Overall | 0.730±0.213 | 0.657±0.090 | 0.671±0.102 | 0.267±0.036 | 0.273±0.042 |
|  | Male | 0.778±0.230 | 0.682±0.079 | 0.699±0.090 | 0.277±0.032 | 0.284±0.036 |
|  | Female | 0.639±0.136 | 0.613±0.092 | 0.620±0.103 | 0.249±0.037 | 0.252±0.042 |
|  | < 30 | 0.749±0.223 | 0.668±0.089 | 0.696±0.102 | 0.271±0.036 | 0.277±0.041 |
|  | ≥30 | 0.711±0.201 | 0.646±0.091 | 0.672±0.105 | 0.263±0.037 | 0.268±0.042 |
| Group 2 | Overall | 0.767±0.217 | - | 0.765±0.183 | - | 0.311±0.075 |
|  | Male | 0.784±0.229 | - | 0.780±0.193 | - | 0.321±0.078 |
|  | Female | 0.700±0.146 | - | 0.662±0.073 | - | 0.269±0.030 |
|  | < 30 | 0.807±0.231 | - | 0.812±0.193 | - | 0.330±0.078 |
|  | ≥30 | 0.662±0.127 | - | 0.636±0.041 | - | 0.259±0.017 |
| Overall | Overall | 0.744±0.215 | 0.657±0.090 | 0.709±0.148 | 0.267±0.036 | 0.288±0.060 |
|  | Male | 0.780±0.229 | 0.682±0.079 | 0.740±0.153 | 0.277±0.032 | 0.301±0.062 |
|  | Female | 0.656±0.141 | 0.613±0.092 | 0.631±0.098 | 0.249±0.037 | 0.257±0.040 |
|  | < 30 | 0.777±0.229 | 0.668±0.089 | 0.746±0.166 | 0.271±0.036 | 0.303±0.067 |
|  | ≥30 | 0.699±0.186 | 0.646±0.091 | 0.653±0.092 | 0.263±0.037 | 0.265±0.037 |

Table 5 The correlation between age, gender, group, and $V_{115-7}$.

|  |  | Age | Gender | Group |
|---|---|---|---|---|
| $V_{115-7}$ (m/s) | Pearson Correlation | -0.263 | 0.319 | -0.312 |
|  | Sig. (1-tailed) | <0.001 | <0.001 | <0.001 |
|  | N | 69 | 69 | 69 |

Note: For each participant, actual age is used; male is represented by the number 1, and female is 0; Group 1 is represented by 1, Group 2 is 0.

**3.2 Comparison with similar experiments**

As listed in Table 6, there are a few experiments on the evacuation of super high-rise building. In comparison to these published experiments, the innovation of

this study is as follows:

1) The highest building of which the evacuation experiment and data have been presented is Shanghai World Financial Center (SWFC), where only 6 participants walk down about 470 m. The experiment in this study is carried out in the second tallest building in the world, and the maximal downward distance that 41 participants travel through is 583 m, which will set a new height record on the evacuation experiment and data.

2) In the experiment carried out in SWFC, only the vertical velocity is presented. The vertical velocity is a useful parameter to estimate directly the evacuation time of super high-rise buildings with different heights. However, the difference of stair structures is not considered in the vertical velocity. For example, to separate the stairwell space between two adjacent refuge floors, there are two common methods, one is changing the location of stairwell (adopted in SWFC), and the other is setting firewall and fire door at the refuge floor (adopted in Shanghai Tower). In the building using former method, pedestrians may have to pass through a long plane distance at the refuge floor, which is not considered in the calculation of vertical velocity. In this study, the detailed structure parameters of the stairwell are obtained, thus the travel distance in the stairwell can be calculated, and then the travel velocity (the local velocity in S125-7 and S115-7) in super high-rise building is presented.

3) Furthermore, with the help of cameras located at different floors, the local velocity in many stair sections are extracted in this study. There are total more than 3700 data points of local velocity, through which the velocity of different genders, ages, and moving distances are classified and analyzed, as shown in Table 4. Moreover, the local velocity in different scenarios can be extracted. For example, as shown in Fig. 6a, the local velocity of Group 1 between the 110$^{th}$ and 100$^{th}$ floor (0.672 ± 0.190 m/s), which is the lowest because of confluence and crowding, can be counted as the velocity in crowding scenario. And then the local velocity of Group 1 between the 40$^{th}$ and 7$^{th}$ floor (0.763 ± 0.228 m/s) can be considered as the velocity in free moving scenario.

4) A new method for calculating density in stairwell is presented in this study, and then the relationship between density and velocity in the evacuation process of super high-rise building is given, as shown in Fig. 11.

Table 6 Comparison with similar experimental data.

| Location | height | number | travel velocity (m/s) | vertical velocity (m/s) |
|---|---|---|---|---|
| A residential building in | 40F | 120 | 0.498 (upward) | - |
| Hong Kong[28] | | | | |
| A residential building in South Korea[29] | 50F (150m) | 30(male) | 0.83 | - |
| | | 30(female) | 0.74 | - |
| SWFC[3] | 101F(470m) | 6 | - | 0.28 |
| Shanghai Tower | 117F(542m) | 28 | 0.771 | 0.310 |
| | 126F(583m) | 41 | 0.735 | 0.267 |
| Shanghai Tower | Crowding | 41 | 0.672 | 0.271 |
| | Free moving | 41 | 0.763 | 0.307 |

## 4 Conclusions

We carried out an evacuation experiment in Shanghai Tower, a 632 m high building with 126 stories. 3 groups (Group 1, 2, and 3) of participants were arranged on different floors (126$^{th}$, 117$^{th}$, and 100$^{th}$) at the beginning and asked to walk downstairs to the ground. The evacuation process in 58 stair sections of 69 participants, 41 in Group 1 and 28 in Group 2, were considered in this study. To detect the movement characteristics in the stairwell of super high-rise buildings, the local velocity was defined here to present the average velocity of each participant traveling through each stair section, and a new measurement method for crowd density in super high-rise building was presented.

The results showed that the evacuation performance in the super high-rise building was related with gender, age, density and vertical distance. First, it was found that pedestrians would set a "suitable velocity" for themselves according to the target height distance before the evacuation, to ensure a relatively stable velocity during the evacuation process, and farther target distance would result in lower "suitable velocity". The evidence was that the Group 1 with a 9.63% higher traveling height spend a 16.39% longer evacuation time, yet within Group 1 or Group 2 the velocity did not decrease with the increase of the moving distance. Second, the density was a key factor in determining whether pedestrians could achieve the "suitable velocity". Specifically, pedestrians couldn't achieve their "suitable velocity" in high density condition, and would speed up to their "suitable velocity" in low density condition. Third, the "suitable velocity" of women was smaller than men in the same scenario, and female "suitable velocity" was more sensitive to the change of the target height distance. At last, younger participants showed better evacuation ability in super high-rise building, including that younger people evacuated faster than the older people, and younger people were less affected by the confluence.

The biggest height distance in the published experiments on the evacuation of super high-rise buildings was about 470 m. And in most of existing studies, only the vertical velocity was presented. In this experiment, the maximal downward height distance that participants travel through was 583 m, which will set a new height record on the evacuation experiment and data. With the help of cameras located at different floors, more than 3700 data points of the travel velocity in many stair sections were extracted, and then the velocity of different genders, ages, and moving distances were classified and analyzed. Furthermore, the relationship between density and velocity in the evacuation process of super high-rise building was given.

The conclusions here are helpful to understand the evacuation behavior rules of people in super high-rise buildings. The experimental results can be used as the basis for setting the speed parameters of people in the emergency evacuation model of super high-rise buildings, and can provide data support for the design and optimization of emergency evacuation capability, and the formulation of emergency evacuation plan for super high-rise buildings.

**Acknowledgements**

This work was supported by Program of Shanghai Science and Technology Committee (19QC1400900).